\documentclass[aps,prb,twocolumn,groupedaddress,showpacs]{revtex4}
\usepackage{amssymb}
\usepackage[fleqn]{amsmath}
\usepackage[dvips]{graphicx}
\usepackage{docs}
\usepackage{bm}
\usepackage{textcomp}
\usepackage[colorlinks=true,linkcolor=blue,citecolor=blue,urlcolor=black]{hyperref}%
\expandafter\ifx\csname package@font\endcsname\relax\else
 \expandafter\expandafter
 \expandafter\usepackage
 \expandafter\expandafter
 \expandafter{\csname package@font\endcsname}%
\fi

\usepackage[papersize={9in,12in}]{geometry}
\newlength{\Figwidth}
\begin{document}
\title{Electrical and ELectronic Properties of Strained Mono-layer InTe}

\author{Shoeib Babaee Touski{\color{blue}$^{1}$}}
\email{touski@hut.ac.ir}
\affiliation{Department of Electrical Engineering, Hamedan University of Technology, Hamedan 65155, Iran}
\author{Mohammad Ariapour{\color{blue}$^{2}$}}
\affiliation{Department of Sciences, Hamedan University of Technology, Hamedan 65155, Iran}
\author{Manouchehr Hosseini{\color{blue}$^{3}$}}\affiliation{{\color{blue}$^{3}$}Department of Electrical Engineering, Bu-Ali Sina University, Hamedan, Iran.}

\begin{abstract}
In this paper, electrical and electronic properties of strained mono-layer InTe for two structures, $\alpha$, and $\beta$ phases, is investigated. The band structure is obtained using density functional theory (DFT). The minimum energy and effective mass of the conduction band and second conduction band for different strains are calculated. A FET with using InTe as the channel material is investigated. Voltage-current characteristics of InTe FET is calculated and I$_{ON}$/I$_{OFF}$ ratio is obtained with respect to biaxial strain.

\end{abstract}



\maketitle

 \section{introduction}
For many years, electronic device dimensions have been scaled down for low-power and high-speed operations but the shrinking dimension reaches severe problem on FET due to short-channel effects (SCEs) \cite{taur2013fundamentals,chaudhry2004controlling}. Although Si has various advantages but to break the scaling limitations, various new materials have been proposed. Ultrathin two-dimensional (2D) semiconductors may show high carrier mobilities for application in small dimension \cite{wang2012electronics,wu2017quantum,li2015boron}. The first member of this family is graphene that many research has done on its application in electronic devices. Due to the gapless nature of graphene, a series of 2D transition metal chalcogenides with well-defined bandgaps, such as MoS$_2$ and WS$_2$ have been proposed \cite{radisavljevic2011single,braga2012quantitative}.

Recently, group-III monochalcogenides (MX, M$=$Ga, In; X$=$S, Se, Te) with honeycomb structures attract great attention due to its layer-by-layer structures and peculiar electronic, optic, and thermal properties \cite{lei2014evolution,mudd2015high,mudd2013tuning}.
They exhibit bandgaps of 1-3 eV\cite{late2012gas} which coincide with the range of visible light with high photo-responsivity \cite{schwarz2014two}. They show high carrier mobility on the range of $\mathrm{10^3 cm^2 /V.s}$ \cite{feng2014back}.


2D InSe, first member of monochalcogenides, has drawn considerable attention due to its the high electron mobility. Mobility is reported as $\mathrm{10^3 cm^2 /V.s}$ for 6-layer InSe \cite{bandurin2017eaves} and mono-layer InSe \cite{kuroda1980resonance} at room temperature. 
Field effect transistors (FET) with 2D InSe as transport channel materials have been reported \cite{feng2014back}. Recently, the mono- and few-layer InSe have been synthesized by mechanical exfoliation \cite{lei2014evolution} and chemical vapor transport methods \cite{ho2016thickness}, which have been widely used in photodetectors with a broadband response from the ultraviolet, visible to the near-infrared region. Photodetectors based on few-layered InSe also show broad spectral responses \cite{tamalampudi2014high,mudd2015high}. The intrinsic electron mobility of 2D InSe up to $\mathrm{10^3 cm^2/V.s}$ and the high current I$_{ON}$/I$_{OFF}$ ratio of $10^8$ in InSe based FETs have been observed in experiments \cite{sucharitakul2015intrinsic}. The 2D InSe and its family are promising candidate for future electronic nano-devices \cite{feng2015performance,feng2014back,sanchez2014electronic}.

Indium Telluride (InTe), another member of III-V monocalgenode, has been theoretically explored and proposed as 2D material with high mobility \cite{zolyomi14}. The band-gap of InTe $(E_{g} = 1.29eV)$ is close to InSe $(\sim 1.37eV)$  \cite{demirci17,zolyomi14}. InTe has been explored for different crystal structure phases \cite{zolyomi2014electrons}. The electronic band structures of InTe mono-layer have been studied with different exchange-correlation potentials. They showed that all of these materials are indirect band-gap semiconductors. It is required to mention that exerting strain is one of the most useful and common ways to tune the electronic and optical properties of these materials \cite{yu2015phase,jalilian17,ariapour19}.

Tuning the electronic and optical properties of nanomaterials help us to find new applications for future electronics. Therefore, in this article, the main goal is the control of the electronic characteristics of InTe by exerting biaxial tensile and compressive strain by using first-principles calculations. This is shown that applying strain can move conduction band minimum (CBM) from $M$-point to $\Gamma$ point \cite{jalilian17}. Strain can decrease band-gap with a linear relation. Strain can change effective mass and tune electronic properties. 

In the next section, we explore computational details. In section \ref{sec:result}, the band structure for different strains is investigated and the minimum energy of the valleys are extracted. The effective mass of the valleys is calculated as a function of strain. In the end, a FET based on mono-layer InTe in the presence of biaxial strain is studied. We present the conclusion of our work in section \ref{sec:conclusion}.

\section{Computational details}
\label{sec:com}
The analysis of mono-layer InTe was carried out using density-functional theory (DFT) as implemented in the Quantum Espresso \cite{giannozzi2009quantum,giannozzi2017advanced} plane-wave-basis codes. To calculate the geometries, band-structures, we used semilocal exchange-correlation functional: PBEsol \cite{perdew2008restoring} functional. The plane-wave cut-off energy was 612 eV. During calculation, a $10\times10$ Monkhorst-Pack k-point grid was used. We were performed full geometry optimization until the forces on the atoms are less than 0.01 eV/̊A. 

We investigate two $\alpha$ and $\beta$ phases of mono-layer InTe. $\alpha$ phase constructs a 2D honeycomb structure, which In and Te pairs vertically placed at two different sub-lattices with $D_{3h}$ group symmetry. The $\beta$ structure shows $D_{3d}$ symmetry which one of the Te layers shifted with respect to the other. This phase breaks the mirror symmetry of the original structure but forms inversion symmetry. The structure of $\alpha$ and $\beta$ phases are explored in our previous work\cite{ariapour19}.  A biaxial strain is applied for each phase and geometry is completely relaxed for two phases. Total energy and lattice constant for two phases are the same. 

The effective mass of electron for each valley is calculated by using the following equation,
\begin{equation}
	m^*=\hbar^2/\left(\partial^2E/\partial k^2\right)
\end{equation}
that $\hbar$ is the reduced Planck constant, E and k are the energy and wave vector at the minimum of the valleys.

\section{results and discussion}
\label{sec:result}

\begin{figure}
	\centering
	\includegraphics[width=0.99\linewidth]{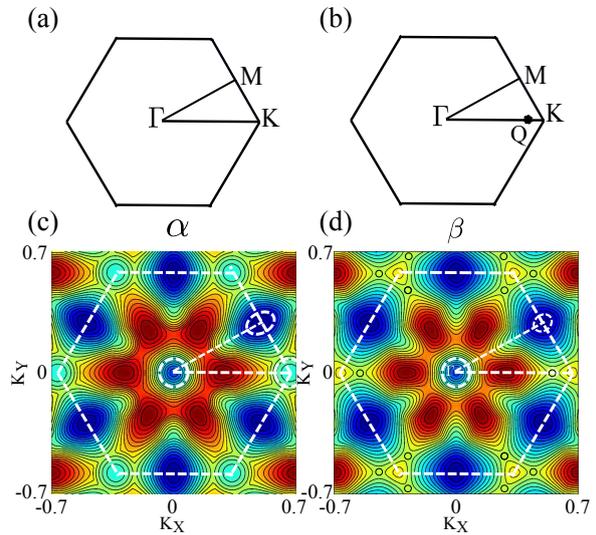}
	\caption{Conduction band energy in the first brilloen zone for (a) $\alpha$ and (b) $\beta$ phases. Three valleys at $\Gamma$, $M$ and $K$ ($Q$) points are specified $\alpha$($\beta$) phase. The path for band structure starts from $\Gamma$ to $K$ and from $K$ to $M$ then returns to $\Gamma$.  }
	\label{fig:fig1}
\end{figure}

InTe has consisted of two In sub-layers is sandwiched between two Te sub-layers. For clarifying of valleys and high symmetry points, conduction band energy in the first Brillouin zone is plotted for $\alpha$ and $\beta$ structures in Fig. \ref{fig:fig1}. $\Gamma$- and $M$-valleys are indicated for both structures. Another valley is located at K-point for $\alpha$ phase whereas, this valley takes some distance from K and is located at Q-point for $\beta$ phase (see Fig. \ref{fig:fig1}). $\Gamma$ and $K$ valleys for $\alpha$ phase and $\Gamma$ and $Q$ valleys for $\beta$ phase show circular contour, whereas, an ellipsoidal contour is obvious for $M$ valley for both structures. An effective mass in the bottom of the conduction band is calculated for $\Gamma$, $K$ and $Q$ valleys. Whereas, two effective masses should be defined for $M$ valley with elliptical shape, longitudinal and transverse effective masses. High symmetry path to plot band structure is indicated for two phases, see Fig. \ref{fig:fig1}. 

\subsection{Conduction Band Minimum}
\begin{figure}
	\centering
	\includegraphics[width=0.95\linewidth]{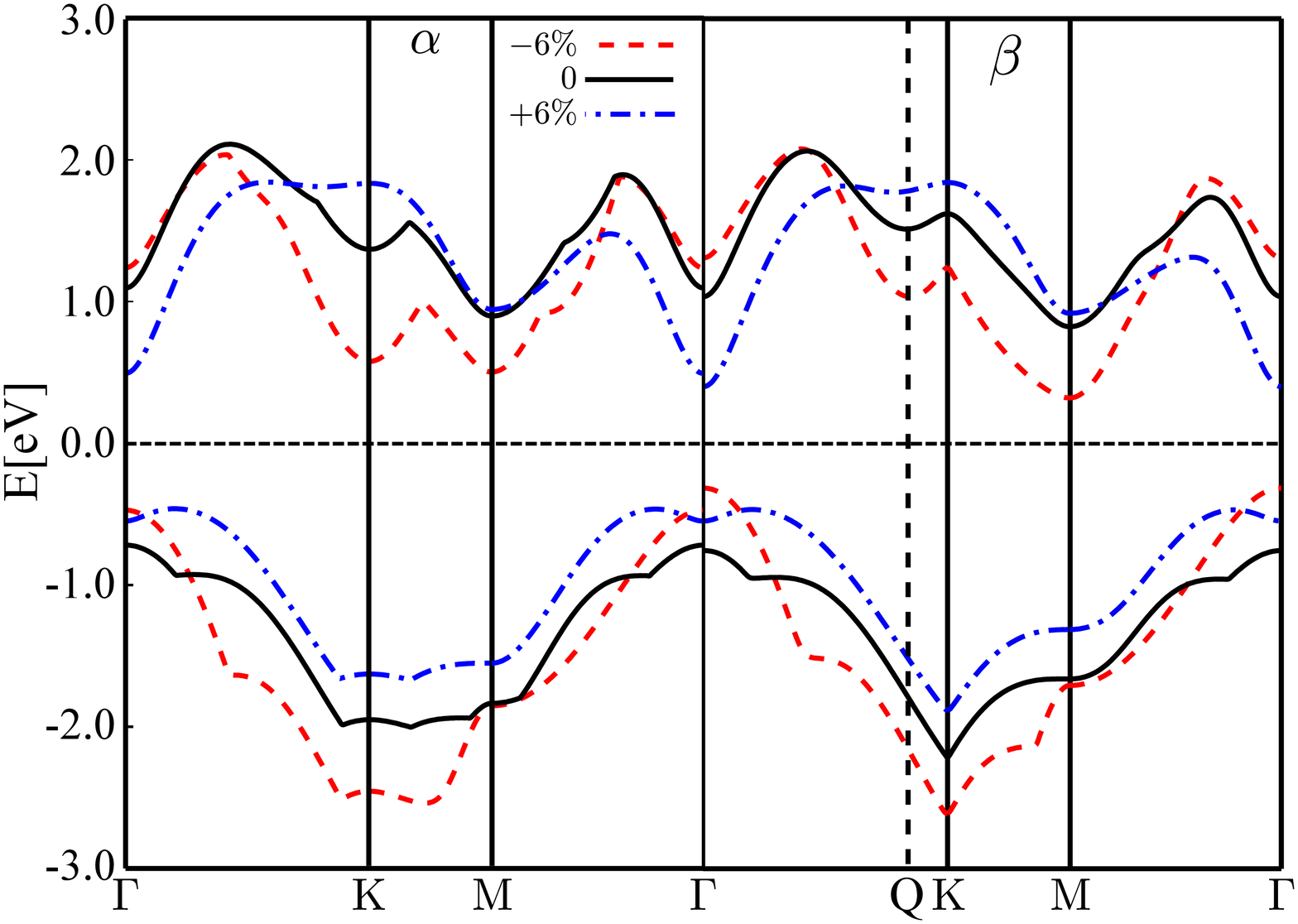}
	\caption{Conduction and valance bands for three states- without strain, compressive and tensile strain- for two $\alpha$ and $\beta$ phases.}
	\label{fig:fig2}
\end{figure}

In Fig. \ref{fig:fig2}, conduction, and valance bands are plotted along the high symmetry paths as is indicated in Fig. \ref{fig:fig1}(a) and (b). The band structure is plotted for three strains- $-6\%$ strain, unstrained and $6\%$ strain. For unstrained one, $M$ valley is lower than other valleys and located at the conduction band minimum (CBM). As one can see, the strain can modify shape and energy for each valley. For example, at $\alpha$ phase with $-6\%$ strain, both $K$ and $M$ valleys are closed to the CBM and contribute to electron transport. In this situation, effective mass for two valleys should be considered. At the other hand, $M$ valley is dominant for $\beta$ phase at $-6\%$ strain. In unstrained one, CBM is located at $M$ valley for both structures. But in $+6\%$  strain, both phases show CBM at the $\Gamma$ valley.

\begin{figure}
	\centering
	\includegraphics[width=0.99\linewidth]{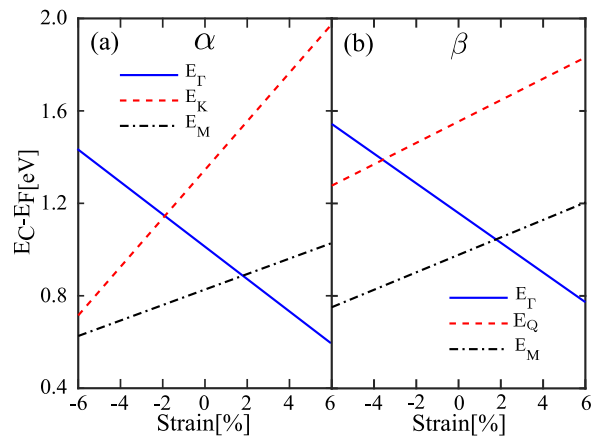}
	\caption{Minimum energy of three valleys versus strain for (a) $\alpha$ and (b) $\beta$ phases. } 
	\label{fig:fig3}
\end{figure}

Strain from $-6\%$ to $+6\%$ is applied to both $\alpha$ and $\beta$ phases and energies at the bottom of the valleys is plotted in Fig. \ref{fig:fig3}. The energy in the bottom of two valleys ($K$ and $M$ for $\alpha$ phase and $Q$ and $M$ for $\beta$ phase) increases respect to strain increasing, whereas, $\Gamma$ valley declines with increases of strain. As one can observe, the minimum energy of valleys change linear respect to strain and an equation is fitted to every curve. Relations between energy and strain form fitted equations are addressed in the following for two $\alpha$ and $\beta$ phases.
\begin{align}
\begin{split}
E^{\alpha}_{\Gamma} &= -0.0702 x + 1.019 \\
E^{\beta}_{\Gamma}  &= -0.0803 x + 0.948 \\
E^{\alpha}_{K}      &=  0.1048 x + 1.301 \\
E^{\beta}_{K}       &=  0.0472 x + 0.72 \\
E^{\alpha}_{M}      &=  0.0335 x + 0.827 \\
E^{\beta}_{M}       &=  0.0578 x + 1.443
\end{split} 
\end{align}
that $x$ is applying strain in percent. $E^{\alpha}$ and $E^{\beta}$ stand for conduction band energies in $\alpha$ and $\beta$ structures, respectively.

Electrons are located at the bottom of the conduction band and CBM highly affects electron transmission. For strains larger than $2\%$, $E^{\alpha}_\Gamma$ is the minimum conduction band and  $E^{\alpha}_M$ is the second minimum conduction band. For strain lower than $2\%$, $E^{\alpha}_M$ comes under other valleys so CBM moves to $M$-valley. $\alpha$ and $\beta$
phases behave similar, but $K$-valley declines in $\alpha$ phase more than it in $\beta$ phase in respect to strain. For strain $-6\%$, $E^{\alpha}_{K}$ reaches close to $E^{\alpha}_{M}$. For both phases, band-gap goes to direct for tensile strain larger than $2\%$. This suggests strained InTe is suitable for optical works at these strains. 
 
\subsection{Effective mass}
\begin{figure}[t]
	\begin{center}
		\includegraphics[width=0.99\linewidth]{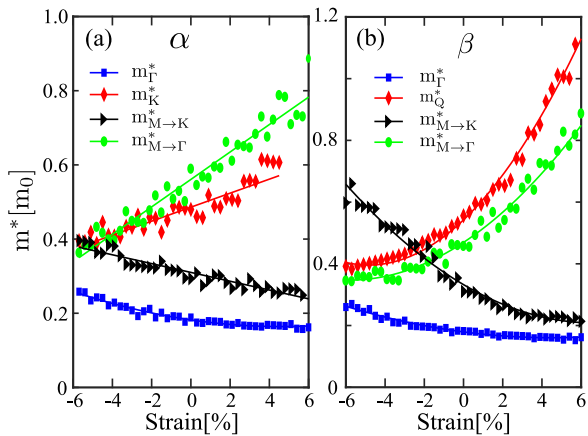}
	\end{center}
	\caption{Effective mass for the valleys as a function of strain for (a) $\alpha$ and (b) $\beta$ phases. } 
	\label{fig:fig4}
\end{figure}

Two valleys on $\Gamma$ and $K$ points, show circular valley that an effective masses are obtained but for the elliptical valley at $M$ point two longitudinal and transverse effective masses are calculated. In Fig. \ref{fig:fig4}, effective mass of the valleys for $\alpha$ and $\beta$ phases are calculated. In unstrained condition, we obtained effective mass for $\Gamma$ equal to 0.19$m_0$ for $\alpha$ phase, that approximately is close to 0.17$m_0$ reported by Ref. [\onlinecite{zolyomi14}]. They reported effective mass for $K$-valley as 0.53$m_0$ that here we obtained 0.47$m_0$.
The longitudinal and transverse effective masses of $M$-valley
are calculated as 0.52$m_0$ and 0.29$m_0$ that are close to reported effective masses \cite{zolyomi14}. These values for the $\beta$ structure are calculated. They reported 0.16$m_0$ for $\Gamma$ valley of $\beta$ phase and we obtained 0.18$m_0$ for it. They didn't report any effective mass for $K$-valley. We observed the valley is located at $Q$ point near to $K$ point.
The effective for this valley is obtained as 0.56$m_0$.  
We calculate the longitudinal and transverse effective masses for $M$-valley as 0.45$m_0$ and 0.31$m_0$.

The Fig. \ref{fig:fig4} shows that $m^{*}_{\Gamma}$ decreases when strain increasing. $\Gamma$-valley shows the smallest effective mass for two phases at all range of strain. The effective mass of strained InTe is about 0.2$m_0$ when CBM has located at $\Gamma$-valley. In the opposite of $\Gamma$-valley, effective mass at $K$-valley for $\alpha$ phase and $Q$-valley for $\beta$ phase increases with strain increasing. For strain larger than $4.5\%$ in $\alpha$ phase, the valley at $K$ point vanishes and no effective mass is reported. For $M$-valley, longitudinal effective mass $m^{*}_{M\rightarrow\Gamma}$ increases with strain, whereas, transverse effective mass $m^{*}_{M\rightarrow K}$ decreases.

Effective masses for both phases smoothly vary with strain. Therefore, a line can be fitted to every effective mass curve. The fitted equations for $\alpha$ phase at $\Gamma$, $K$, $M\rightarrow\Gamma$ and $M\rightarrow K$ respect with strain $(x)$ are listed in the following.
\begin{align}
\begin{split}
&m^{*\alpha}_{\Gamma} = 0.00084x^2-0.007557 x+0.182 \\
&m^{*\alpha}_{K} = 0.01889 x + 0.486 \\
&m^{*\alpha}_{M\rightarrow \Gamma} = 0.037 x + 0.5608 \\
&m^{*\alpha}_{M\rightarrow K} = -0.01188 x + 0.3101 
\label{msa}
\end{split}
\end{align}
In the following, equations for fitted line of effective mass respect to strain $(x)$ in $\beta$ phase are as:
\begin{align}
\begin{split}
&m^{*\beta}_{\Gamma} = 0.0009206 x^{2} - 0.008329 x +0.1793 \\
&m^{*\beta}_{Q} = 0.00624x^2 + 0.06043 x + 0.5422 \\
&m^{*\beta}_{M\rightarrow \Gamma} = 0.00357x^2 + 0.04179x + 0.4693\\
&m^{*\beta}_{M\rightarrow K} = 0.002808 x^{2} -0.03733 x + 0.3325 
\label{msb}
\end{split}
\end{align} 
Effective mass at $\Gamma$-valley for two phases show similar variation respect to strain. One can conclude two phases behave similarly when CBM is located at $\Gamma$-valley.

\begin{figure}
	\centering
	\includegraphics[width=1.0\linewidth]{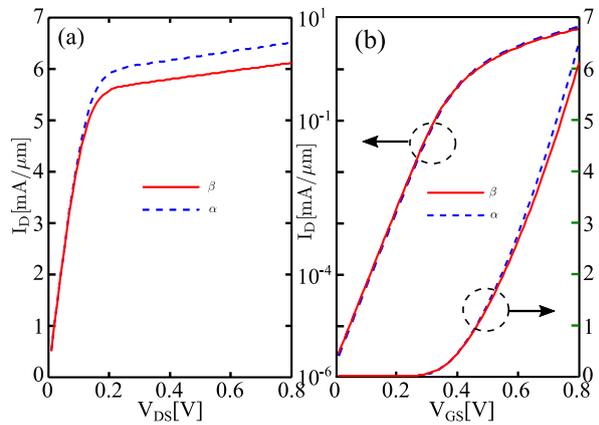}	\caption{ Drain current versus (a) $V_{DS}$ and (b) $V_{GS}$ for two $\alpha$ and $\beta$ phases. }
	\label{fig:fig5}
\end{figure}

\subsection{FET results}
In this section, the electronic properties of strained InTe are studied. Mono-layer of InTe is used as a channel in a FET. Drain current is calculated by using the top of the barrier model. $I_{DS}$ as a function of drain voltage for two phases is plotted in Fig. \ref{fig:fig5}(a). $I_{DS}$ increases with $V_{DS}$ up to $\mathrm{0.2V}$ in linear regime then is saturated. Therefore, $\mathrm{V_{DS}=0.2V}$ is enough to bring in the saturation regime. $\alpha$ phase with lower effective mass shows higher current in the saturated regime. 

$I_{DS}$ as a function of gate voltage in linear and logarithmic scale is plotted in Fig. \ref{fig:fig5}(b). $I_{DS}$ increases exponentially with respect to the gate voltage in the sub-threshold regime and approximately are saturated for $\mathrm{V_{GS}=0.35V}$. Two phases behave similarly in logarithmic scale whereas $\alpha$ phase indicates higher ON-current ($I_{ON}$) in linear scale. 

\begin{figure}[t]
	\centering
	\includegraphics[width=0.8\linewidth]{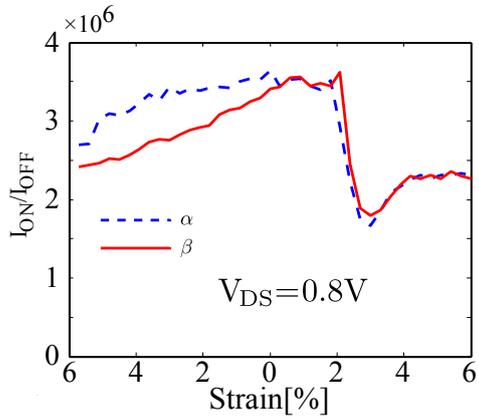}
	\caption{ON-OFF ratio as a function of strain for $\alpha$ and $\beta$ phases.}
	\label{fig:ionioff}
\end{figure}

I$_{ON}$/I$_{OFF}$ ratio, an important parameter of a FET, is plotted as a function of strain for two phases in Fig. \ref{fig:ionioff}. This parameter stands larger than $10^6$ for all range of strain. I$_{ON}$/I$_{OFF}$ ratio increase respect to strain up to $2\%$ then falls down. CBM is located at M-valley for strain in the interval [-6$\%$,2$\%$] and goes to $\Gamma$-valley for strain larger than $2\%$. 
Changing CBM from M-valley to $\Gamma$-valley causes I$_{ON}$/I$_{OFF}$ ratio falls down. Although, $\Gamma$-valley shows less effective mass but six valleys for M-valley in the first  Brillouin zone contribute to larger ON-current and larger I$_{ON}$/I$_{OFF}$. For strain in the interval [-6$\%$,2$\%$], effective mass at M-valley decreases and ON-current increment causes I$_{ON}$/I$_{OFF}$ ratio increases.
The results indicate that tensile strain decreases the performance of InTe FET. I$_{ON}$/I$_{OFF}$ ratio can decrease two times by tensile strain.

\section{conclusions}
\label{sec:conclusion}
Location of the valleys in the first Brillouin zone for mono-layer InTe is obtained. Three valleys at M, K(Q) and $\Gamma$ points for $\alpha$($\beta$) phase contribute to electrical properties. The minimum energy of these valleys as a function of strain is calculated. It is observed CBM is located at $\Gamma$-point for strain larger than $2\%$ whereas lies at $M$-valley for smaller strain. Effective mass as a function of strain is calculated for these valleys and a curve is fitted to effective mass and minimum energy of each valley. $\Gamma$-valley shows the lowest effective mass for all the range of strain.  The results for FET simulation indicates an I$_{ON}$/I$_{OFF}$ ratio in the range $10^6$ and tensile strain decreases it until two times.  


\begin{thebibliography}{29}
	\expandafter\ifx\csname natexlab\endcsname\relax\def\natexlab#1{#1}\fi
	\expandafter\ifx\csname bibnamefont\endcsname\relax
	\def\bibnamefont#1{#1}\fi
	\expandafter\ifx\csname bibfnamefont\endcsname\relax
	\def\bibfnamefont#1{#1}\fi
	\expandafter\ifx\csname citenamefont\endcsname\relax
	\def\citenamefont#1{#1}\fi
	\expandafter\ifx\csname url\endcsname\relax
	\def\url#1{\texttt{#1}}\fi
	\expandafter\ifx\csname urlprefix\endcsname\relax\def\urlprefix{URL }\fi
	\providecommand{\bibinfo}[2]{#2}
	\providecommand{\eprint}[2][]{\url{#2}}
	
	\bibitem[{\citenamefont{Taur and Ning}(2013)}]{taur2013fundamentals}
	\bibinfo{author}{\bibfnamefont{Y.}~\bibnamefont{Taur}} \bibnamefont{and}
	\bibinfo{author}{\bibfnamefont{T.~H.} \bibnamefont{Ning}},
	\emph{\bibinfo{title}{Fundamentals of modern VLSI devices}}
	(\bibinfo{publisher}{Cambridge university press}, \bibinfo{year}{2013}).
	
	\bibitem[{\citenamefont{Chaudhry and Kumar}(2004)}]{chaudhry2004controlling}
	\bibinfo{author}{\bibfnamefont{A.}~\bibnamefont{Chaudhry}} \bibnamefont{and}
	\bibinfo{author}{\bibfnamefont{M.~J.} \bibnamefont{Kumar}},
	\bibinfo{journal}{IEEE Trans. Device Mater. Reliab.}
	\textbf{\bibinfo{volume}{4}}, \bibinfo{pages}{99} (\bibinfo{year}{2004}).
	
	\bibitem[{\citenamefont{Wang et~al.}(2012)\citenamefont{Wang, Kalantar-Zadeh,
			Kis, Coleman, and Strano}}]{wang2012electronics}
	\bibinfo{author}{\bibfnamefont{Q.~H.} \bibnamefont{Wang}},
	\bibinfo{author}{\bibfnamefont{K.}~\bibnamefont{Kalantar-Zadeh}},
	\bibinfo{author}{\bibfnamefont{A.}~\bibnamefont{Kis}},
	\bibinfo{author}{\bibfnamefont{J.~N.} \bibnamefont{Coleman}},
	\bibnamefont{and} \bibinfo{author}{\bibfnamefont{M.~S.}
		\bibnamefont{Strano}}, \bibinfo{journal}{Nat. Nanotechnol.}
	\textbf{\bibinfo{volume}{7}}, \bibinfo{pages}{699} (\bibinfo{year}{2012}).
	
	\bibitem[{\citenamefont{Wu et~al.}(2017)\citenamefont{Wu, Zhang, Lee, Duesberg,
			Syrlybekov, Liu, Abid, Abid, Liu, Zhang et~al.}}]{wu2017quantum}
	\bibinfo{author}{\bibfnamefont{Y.}~\bibnamefont{Wu}},
	\bibinfo{author}{\bibfnamefont{D.}~\bibnamefont{Zhang}},
	\bibinfo{author}{\bibfnamefont{K.}~\bibnamefont{Lee}},
	\bibinfo{author}{\bibfnamefont{G.~S.} \bibnamefont{Duesberg}},
	\bibinfo{author}{\bibfnamefont{A.}~\bibnamefont{Syrlybekov}},
	\bibinfo{author}{\bibfnamefont{X.}~\bibnamefont{Liu}},
	\bibinfo{author}{\bibfnamefont{M.}~\bibnamefont{Abid}},
	\bibinfo{author}{\bibfnamefont{M.}~\bibnamefont{Abid}},
	\bibinfo{author}{\bibfnamefont{Y.}~\bibnamefont{Liu}},
	\bibinfo{author}{\bibfnamefont{L.}~\bibnamefont{Zhang}},
	\bibnamefont{et~al.}, \bibinfo{journal}{Adv. Mater. Technol.}
	\textbf{\bibinfo{volume}{2}}, \bibinfo{pages}{1600197}
	(\bibinfo{year}{2017}).
	
	\bibitem[{\citenamefont{Li et~al.}(2015)\citenamefont{Li, Xie, Zheng, Tian, and
			Sun}}]{li2015boron}
	\bibinfo{author}{\bibfnamefont{X.-B.} \bibnamefont{Li}},
	\bibinfo{author}{\bibfnamefont{S.-Y.} \bibnamefont{Xie}},
	\bibinfo{author}{\bibfnamefont{H.}~\bibnamefont{Zheng}},
	\bibinfo{author}{\bibfnamefont{W.~Q.} \bibnamefont{Tian}}, \bibnamefont{and}
	\bibinfo{author}{\bibfnamefont{H.-B.} \bibnamefont{Sun}},
	\bibinfo{journal}{Nanoscale} \textbf{\bibinfo{volume}{7}},
	\bibinfo{pages}{18863} (\bibinfo{year}{2015}).
	
	\bibitem[{\citenamefont{Radisavljevic et~al.}(2011)\citenamefont{Radisavljevic,
			Radenovic, Brivio, Giacometti, and Kis}}]{radisavljevic2011single}
	\bibinfo{author}{\bibfnamefont{B.}~\bibnamefont{Radisavljevic}},
	\bibinfo{author}{\bibfnamefont{A.}~\bibnamefont{Radenovic}},
	\bibinfo{author}{\bibfnamefont{J.}~\bibnamefont{Brivio}},
	\bibinfo{author}{\bibfnamefont{i.~V.} \bibnamefont{Giacometti}},
	\bibnamefont{and} \bibinfo{author}{\bibfnamefont{A.}~\bibnamefont{Kis}},
	\bibinfo{journal}{Nat. Nanotechnol.} \textbf{\bibinfo{volume}{6}},
	\bibinfo{pages}{147} (\bibinfo{year}{2011}).
	
	\bibitem[{\citenamefont{Braga et~al.}(2012)\citenamefont{Braga,
			Gutiérrez~Lezama, Berger, and Morpurgo}}]{braga2012quantitative}
	\bibinfo{author}{\bibfnamefont{D.}~\bibnamefont{Braga}},
	\bibinfo{author}{\bibfnamefont{I.}~\bibnamefont{Gutiérrez~Lezama}},
	\bibinfo{author}{\bibfnamefont{H.}~\bibnamefont{Berger}}, \bibnamefont{and}
	\bibinfo{author}{\bibfnamefont{A.~F.} \bibnamefont{Morpurgo}},
	\bibinfo{journal}{Nano Lett.} \textbf{\bibinfo{volume}{12}},
	\bibinfo{pages}{5218} (\bibinfo{year}{2012}).
	
	\bibitem[{\citenamefont{Lei et~al.}(2014)\citenamefont{Lei, Ge, Najmaei,
			George, Kappera, Lou, Chhowalla, Yamaguchi, Gupta, Vajtai
			et~al.}}]{lei2014evolution}
	\bibinfo{author}{\bibfnamefont{S.}~\bibnamefont{Lei}},
	\bibinfo{author}{\bibfnamefont{L.}~\bibnamefont{Ge}},
	\bibinfo{author}{\bibfnamefont{S.}~\bibnamefont{Najmaei}},
	\bibinfo{author}{\bibfnamefont{A.}~\bibnamefont{George}},
	\bibinfo{author}{\bibfnamefont{R.}~\bibnamefont{Kappera}},
	\bibinfo{author}{\bibfnamefont{J.}~\bibnamefont{Lou}},
	\bibinfo{author}{\bibfnamefont{M.}~\bibnamefont{Chhowalla}},
	\bibinfo{author}{\bibfnamefont{H.}~\bibnamefont{Yamaguchi}},
	\bibinfo{author}{\bibfnamefont{G.}~\bibnamefont{Gupta}},
	\bibinfo{author}{\bibfnamefont{R.}~\bibnamefont{Vajtai}},
	\bibnamefont{et~al.}, \bibinfo{journal}{ACS nano}
	\textbf{\bibinfo{volume}{8}}, \bibinfo{pages}{1263} (\bibinfo{year}{2014}).
	
	\bibitem[{\citenamefont{Mudd et~al.}(2015)\citenamefont{Mudd, Svatek, Hague,
			Makarovsky, Kudrynskyi, Mellor, Beton, Eaves, Novoselov, Kovalyuk
			et~al.}}]{mudd2015high}
	\bibinfo{author}{\bibfnamefont{G.~W.} \bibnamefont{Mudd}},
	\bibinfo{author}{\bibfnamefont{S.~A.} \bibnamefont{Svatek}},
	\bibinfo{author}{\bibfnamefont{L.}~\bibnamefont{Hague}},
	\bibinfo{author}{\bibfnamefont{O.}~\bibnamefont{Makarovsky}},
	\bibinfo{author}{\bibfnamefont{Z.~R.} \bibnamefont{Kudrynskyi}},
	\bibinfo{author}{\bibfnamefont{C.~J.} \bibnamefont{Mellor}},
	\bibinfo{author}{\bibfnamefont{P.~H.} \bibnamefont{Beton}},
	\bibinfo{author}{\bibfnamefont{L.}~\bibnamefont{Eaves}},
	\bibinfo{author}{\bibfnamefont{K.~S.} \bibnamefont{Novoselov}},
	\bibinfo{author}{\bibfnamefont{Z.~D.} \bibnamefont{Kovalyuk}},
	\bibnamefont{et~al.}, \bibinfo{journal}{Adv. Mater.}
	\textbf{\bibinfo{volume}{27}}, \bibinfo{pages}{3760} (\bibinfo{year}{2015}).
	
	\bibitem[{\citenamefont{Mudd et~al.}(2013)\citenamefont{Mudd, Svatek, Ren,
			Patan{\`e}, Makarovsky, Eaves, Beton, Kovalyuk, Lashkarev, Kudrynskyi
			et~al.}}]{mudd2013tuning}
	\bibinfo{author}{\bibfnamefont{G.~W.} \bibnamefont{Mudd}},
	\bibinfo{author}{\bibfnamefont{S.~A.} \bibnamefont{Svatek}},
	\bibinfo{author}{\bibfnamefont{T.}~\bibnamefont{Ren}},
	\bibinfo{author}{\bibfnamefont{A.}~\bibnamefont{Patan{\`e}}},
	\bibinfo{author}{\bibfnamefont{O.}~\bibnamefont{Makarovsky}},
	\bibinfo{author}{\bibfnamefont{L.}~\bibnamefont{Eaves}},
	\bibinfo{author}{\bibfnamefont{P.~H.} \bibnamefont{Beton}},
	\bibinfo{author}{\bibfnamefont{Z.~D.} \bibnamefont{Kovalyuk}},
	\bibinfo{author}{\bibfnamefont{G.~V.} \bibnamefont{Lashkarev}},
	\bibinfo{author}{\bibfnamefont{Z.~R.} \bibnamefont{Kudrynskyi}},
	\bibnamefont{et~al.}, \bibinfo{journal}{Adv. Mater.}
	\textbf{\bibinfo{volume}{25}}, \bibinfo{pages}{5714} (\bibinfo{year}{2013}).
	
	\bibitem[{\citenamefont{Late et~al.}(2012)\citenamefont{Late, Liu, Luo, Yan,
			Matte, Grayson, Rao, and Dravid}}]{late2012gas}
	\bibinfo{author}{\bibfnamefont{D.~J.} \bibnamefont{Late}},
	\bibinfo{author}{\bibfnamefont{B.}~\bibnamefont{Liu}},
	\bibinfo{author}{\bibfnamefont{J.}~\bibnamefont{Luo}},
	\bibinfo{author}{\bibfnamefont{A.}~\bibnamefont{Yan}},
	\bibinfo{author}{\bibfnamefont{H.~R.} \bibnamefont{Matte}},
	\bibinfo{author}{\bibfnamefont{M.}~\bibnamefont{Grayson}},
	\bibinfo{author}{\bibfnamefont{C.}~\bibnamefont{Rao}}, \bibnamefont{and}
	\bibinfo{author}{\bibfnamefont{V.~P.} \bibnamefont{Dravid}},
	\bibinfo{journal}{Adv. Mater.} \textbf{\bibinfo{volume}{24}},
	\bibinfo{pages}{3549} (\bibinfo{year}{2012}).
	
	\bibitem[{\citenamefont{Schwarz et~al.}(2014)\citenamefont{Schwarz, Dufferwiel,
			Walker, Withers, Trichet, Sich, Li, Chekhovich, Borisenko, Kolesnikov
			et~al.}}]{schwarz2014two}
	\bibinfo{author}{\bibfnamefont{S.}~\bibnamefont{Schwarz}},
	\bibinfo{author}{\bibfnamefont{S.}~\bibnamefont{Dufferwiel}},
	\bibinfo{author}{\bibfnamefont{P.}~\bibnamefont{Walker}},
	\bibinfo{author}{\bibfnamefont{F.}~\bibnamefont{Withers}},
	\bibinfo{author}{\bibfnamefont{A.}~\bibnamefont{Trichet}},
	\bibinfo{author}{\bibfnamefont{M.}~\bibnamefont{Sich}},
	\bibinfo{author}{\bibfnamefont{F.}~\bibnamefont{Li}},
	\bibinfo{author}{\bibfnamefont{E.}~\bibnamefont{Chekhovich}},
	\bibinfo{author}{\bibfnamefont{D.}~\bibnamefont{Borisenko}},
	\bibinfo{author}{\bibfnamefont{N.~N.} \bibnamefont{Kolesnikov}},
	\bibnamefont{et~al.}, \bibinfo{journal}{Nano Lett.}
	\textbf{\bibinfo{volume}{14}}, \bibinfo{pages}{7003} (\bibinfo{year}{2014}).
	
	\bibitem[{\citenamefont{Feng et~al.}(2014)\citenamefont{Feng, Zheng, Cao, and
			Hu}}]{feng2014back}
	\bibinfo{author}{\bibfnamefont{W.}~\bibnamefont{Feng}},
	\bibinfo{author}{\bibfnamefont{W.}~\bibnamefont{Zheng}},
	\bibinfo{author}{\bibfnamefont{W.}~\bibnamefont{Cao}}, \bibnamefont{and}
	\bibinfo{author}{\bibfnamefont{P.}~\bibnamefont{Hu}},
	\bibinfo{journal}{Advanced Materials} \textbf{\bibinfo{volume}{26}},
	\bibinfo{pages}{6587} (\bibinfo{year}{2014}).
	
	\bibitem[{\citenamefont{Bandurin et~al.}(2017)\citenamefont{Bandurin, Tyurnina,
			Yu, Mishchenko, Z{\'o}lyomi, Morozov, Kumar, Gorbachev, Kudrynskyi, Pezzini
			et~al.}}]{bandurin2017eaves}
	\bibinfo{author}{\bibfnamefont{D.}~\bibnamefont{Bandurin}},
	\bibinfo{author}{\bibfnamefont{A.}~\bibnamefont{Tyurnina}},
	\bibinfo{author}{\bibfnamefont{G.}~\bibnamefont{Yu}},
	\bibinfo{author}{\bibfnamefont{A.}~\bibnamefont{Mishchenko}},
	\bibinfo{author}{\bibfnamefont{V.}~\bibnamefont{Z{\'o}lyomi}},
	\bibinfo{author}{\bibfnamefont{S.}~\bibnamefont{Morozov}},
	\bibinfo{author}{\bibfnamefont{R.~K.} \bibnamefont{Kumar}},
	\bibinfo{author}{\bibfnamefont{R.}~\bibnamefont{Gorbachev}},
	\bibinfo{author}{\bibfnamefont{Z.}~\bibnamefont{Kudrynskyi}},
	\bibinfo{author}{\bibfnamefont{S.}~\bibnamefont{Pezzini}},
	\bibnamefont{et~al.}, \bibinfo{journal}{Nat. Nanotechnol.}
	\textbf{\bibinfo{volume}{12}}, \bibinfo{pages}{223} (\bibinfo{year}{2017}).
	
	\bibitem[{\citenamefont{Kuroda and Nishina}(1980)}]{kuroda1980resonance}
	\bibinfo{author}{\bibfnamefont{N.}~\bibnamefont{Kuroda}} \bibnamefont{and}
	\bibinfo{author}{\bibfnamefont{Y.}~\bibnamefont{Nishina}},
	\bibinfo{journal}{Solid State Commun.} \textbf{\bibinfo{volume}{34}},
	\bibinfo{pages}{481} (\bibinfo{year}{1980}).
	
	\bibitem[{\citenamefont{Ho}(2016)}]{ho2016thickness}
	\bibinfo{author}{\bibfnamefont{C.-H.} \bibnamefont{Ho}}, \bibinfo{journal}{2D
		Mater.} \textbf{\bibinfo{volume}{3}}, \bibinfo{pages}{025019}
	(\bibinfo{year}{2016}).
	
	\bibitem[{\citenamefont{Tamalampudi et~al.}(2014)\citenamefont{Tamalampudi, Lu,
			Kumar~U, Sankar, Liao, Moorthy~B, Cheng, Chou, and
			Chen}}]{tamalampudi2014high}
	\bibinfo{author}{\bibfnamefont{S.~R.} \bibnamefont{Tamalampudi}},
	\bibinfo{author}{\bibfnamefont{Y.-Y.} \bibnamefont{Lu}},
	\bibinfo{author}{\bibfnamefont{R.}~\bibnamefont{Kumar~U}},
	\bibinfo{author}{\bibfnamefont{R.}~\bibnamefont{Sankar}},
	\bibinfo{author}{\bibfnamefont{C.-D.} \bibnamefont{Liao}},
	\bibinfo{author}{\bibfnamefont{K.}~\bibnamefont{Moorthy~B}},
	\bibinfo{author}{\bibfnamefont{C.-H.} \bibnamefont{Cheng}},
	\bibinfo{author}{\bibfnamefont{F.~C.} \bibnamefont{Chou}}, \bibnamefont{and}
	\bibinfo{author}{\bibfnamefont{Y.-T.} \bibnamefont{Chen}},
	\bibinfo{journal}{Nano Lett.} \textbf{\bibinfo{volume}{14}},
	\bibinfo{pages}{2800} (\bibinfo{year}{2014}).
	
	\bibitem[{\citenamefont{Sucharitakul et~al.}(2015)\citenamefont{Sucharitakul,
			Goble, Kumar, Sankar, Bogorad, Chou, Chen, and
			Gao}}]{sucharitakul2015intrinsic}
	\bibinfo{author}{\bibfnamefont{S.}~\bibnamefont{Sucharitakul}},
	\bibinfo{author}{\bibfnamefont{N.~J.} \bibnamefont{Goble}},
	\bibinfo{author}{\bibfnamefont{U.~R.} \bibnamefont{Kumar}},
	\bibinfo{author}{\bibfnamefont{R.}~\bibnamefont{Sankar}},
	\bibinfo{author}{\bibfnamefont{Z.~A.} \bibnamefont{Bogorad}},
	\bibinfo{author}{\bibfnamefont{F.-C.} \bibnamefont{Chou}},
	\bibinfo{author}{\bibfnamefont{Y.-T.} \bibnamefont{Chen}}, \bibnamefont{and}
	\bibinfo{author}{\bibfnamefont{X.~P.} \bibnamefont{Gao}},
	\bibinfo{journal}{Nano Lett.} \textbf{\bibinfo{volume}{15}},
	\bibinfo{pages}{3815} (\bibinfo{year}{2015}).
	
	\bibitem[{\citenamefont{Feng et~al.}(2015)\citenamefont{Feng, Zhou, Tian,
			Zheng, and Hu}}]{feng2015performance}
	\bibinfo{author}{\bibfnamefont{W.}~\bibnamefont{Feng}},
	\bibinfo{author}{\bibfnamefont{X.}~\bibnamefont{Zhou}},
	\bibinfo{author}{\bibfnamefont{W.~Q.} \bibnamefont{Tian}},
	\bibinfo{author}{\bibfnamefont{W.}~\bibnamefont{Zheng}}, \bibnamefont{and}
	\bibinfo{author}{\bibfnamefont{P.}~\bibnamefont{Hu}}, \bibinfo{journal}{Phys.
		Chem. Chem. Phys.} \textbf{\bibinfo{volume}{17}}, \bibinfo{pages}{3653}
	(\bibinfo{year}{2015}).
	
	\bibitem[{\citenamefont{S{\'a}nchez-Royo
			et~al.}(2014)\citenamefont{S{\'a}nchez-Royo, Mu{\~n}oz-Matutano,
			Brotons-Gisbert, Mart{\'\i}nez-Pastor, Segura, Cantarero, Mata, Canet-Ferrer,
			Tobias, Canadell et~al.}}]{sanchez2014electronic}
	\bibinfo{author}{\bibfnamefont{J.~F.} \bibnamefont{S{\'a}nchez-Royo}},
	\bibinfo{author}{\bibfnamefont{G.}~\bibnamefont{Mu{\~n}oz-Matutano}},
	\bibinfo{author}{\bibfnamefont{M.}~\bibnamefont{Brotons-Gisbert}},
	\bibinfo{author}{\bibfnamefont{J.~P.} \bibnamefont{Mart{\'\i}nez-Pastor}},
	\bibinfo{author}{\bibfnamefont{A.}~\bibnamefont{Segura}},
	\bibinfo{author}{\bibfnamefont{A.}~\bibnamefont{Cantarero}},
	\bibinfo{author}{\bibfnamefont{R.}~\bibnamefont{Mata}},
	\bibinfo{author}{\bibfnamefont{J.}~\bibnamefont{Canet-Ferrer}},
	\bibinfo{author}{\bibfnamefont{G.}~\bibnamefont{Tobias}},
	\bibinfo{author}{\bibfnamefont{E.}~\bibnamefont{Canadell}},
	\bibnamefont{et~al.}, \bibinfo{journal}{Nano Res.}
	\textbf{\bibinfo{volume}{7}}, \bibinfo{pages}{1556} (\bibinfo{year}{2014}).
	
	\bibitem[{\citenamefont{Z{\'o}lyomi
			et~al.}(2014{\natexlab{a}})\citenamefont{Z{\'o}lyomi, Drummond, and
			Fal'Ko}}]{zolyomi14}
	\bibinfo{author}{\bibfnamefont{V.}~\bibnamefont{Z{\'o}lyomi}},
	\bibinfo{author}{\bibfnamefont{N.}~\bibnamefont{Drummond}}, \bibnamefont{and}
	\bibinfo{author}{\bibfnamefont{V.}~\bibnamefont{Fal'Ko}},
	\bibinfo{journal}{Phys. Rev. B} \textbf{\bibinfo{volume}{89}},
	\bibinfo{pages}{205416} (\bibinfo{year}{2014}{\natexlab{a}}).
	
	\bibitem[{\citenamefont{Demirci et~al.}(2017)\citenamefont{Demirci, Avazl{\i},
			Durgun, and Cahangirov}}]{demirci17}
	\bibinfo{author}{\bibfnamefont{S.}~\bibnamefont{Demirci}},
	\bibinfo{author}{\bibfnamefont{N.}~\bibnamefont{Avazl{\i}}},
	\bibinfo{author}{\bibfnamefont{E.}~\bibnamefont{Durgun}}, \bibnamefont{and}
	\bibinfo{author}{\bibfnamefont{S.}~\bibnamefont{Cahangirov}},
	\bibinfo{journal}{Phys. Rev. B} \textbf{\bibinfo{volume}{95}},
	\bibinfo{pages}{115409} (\bibinfo{year}{2017}).
	
	\bibitem[{\citenamefont{Z{\'o}lyomi
			et~al.}(2014{\natexlab{b}})\citenamefont{Z{\'o}lyomi, Drummond, and
			Fal'Ko}}]{zolyomi2014electrons}
	\bibinfo{author}{\bibfnamefont{V.}~\bibnamefont{Z{\'o}lyomi}},
	\bibinfo{author}{\bibfnamefont{N.}~\bibnamefont{Drummond}}, \bibnamefont{and}
	\bibinfo{author}{\bibfnamefont{V.}~\bibnamefont{Fal'Ko}},
	\bibinfo{journal}{Phys. Rev. B} \textbf{\bibinfo{volume}{89}},
	\bibinfo{pages}{205416} (\bibinfo{year}{2014}{\natexlab{b}}).
	
	\bibitem[{\citenamefont{Yu et~al.}(2015)\citenamefont{Yu, Xiong, Eshun, Yuan,
			and Li}}]{yu2015phase}
	\bibinfo{author}{\bibfnamefont{S.}~\bibnamefont{Yu}},
	\bibinfo{author}{\bibfnamefont{H.~D.} \bibnamefont{Xiong}},
	\bibinfo{author}{\bibfnamefont{K.}~\bibnamefont{Eshun}},
	\bibinfo{author}{\bibfnamefont{H.}~\bibnamefont{Yuan}}, \bibnamefont{and}
	\bibinfo{author}{\bibfnamefont{Q.}~\bibnamefont{Li}}, \bibinfo{journal}{Appl.
		Surf. Sci.} \textbf{\bibinfo{volume}{325}}, \bibinfo{pages}{27}
	(\bibinfo{year}{2015}).
	
	\bibitem[{\citenamefont{Jalilian and Safari}(2017)}]{jalilian17}
	\bibinfo{author}{\bibfnamefont{J.}~\bibnamefont{Jalilian}} \bibnamefont{and}
	\bibinfo{author}{\bibfnamefont{M.}~\bibnamefont{Safari}},
	\bibinfo{journal}{Phys. Lett. A} \textbf{\bibinfo{volume}{381}},
	\bibinfo{pages}{1313} (\bibinfo{year}{2017}).
	
	\bibitem[{\citenamefont{Ariapour and Touski}(2019)}]{ariapour19}
	\bibinfo{author}{\bibfnamefont{M.}~\bibnamefont{Ariapour}} \bibnamefont{and}
	\bibinfo{author}{\bibfnamefont{S.~B.} \bibnamefont{Touski}},
	\bibinfo{journal}{Mater. Res. Express} \textbf{\bibinfo{volume}{6}},
	\bibinfo{pages}{076402} (\bibinfo{year}{2019}).
	
	\bibitem[{\citenamefont{Giannozzi et~al.}(2009)\citenamefont{Giannozzi, Baroni,
			Bonini, Calandra, Car, Cavazzoni, Ceresoli, Chiarotti, Cococcioni, Dabo
			et~al.}}]{giannozzi2009quantum}
	\bibinfo{author}{\bibfnamefont{P.}~\bibnamefont{Giannozzi}},
	\bibinfo{author}{\bibfnamefont{S.}~\bibnamefont{Baroni}},
	\bibinfo{author}{\bibfnamefont{N.}~\bibnamefont{Bonini}},
	\bibinfo{author}{\bibfnamefont{M.}~\bibnamefont{Calandra}},
	\bibinfo{author}{\bibfnamefont{R.}~\bibnamefont{Car}},
	\bibinfo{author}{\bibfnamefont{C.}~\bibnamefont{Cavazzoni}},
	\bibinfo{author}{\bibfnamefont{D.}~\bibnamefont{Ceresoli}},
	\bibinfo{author}{\bibfnamefont{G.~L.} \bibnamefont{Chiarotti}},
	\bibinfo{author}{\bibfnamefont{M.}~\bibnamefont{Cococcioni}},
	\bibinfo{author}{\bibfnamefont{I.}~\bibnamefont{Dabo}}, \bibnamefont{et~al.},
	\bibinfo{journal}{J. Phys.:Condens. Matter} \textbf{\bibinfo{volume}{21}},
	\bibinfo{pages}{395502} (\bibinfo{year}{2009}).
	
	\bibitem[{\citenamefont{Giannozzi et~al.}(2017)\citenamefont{Giannozzi,
			Andreussi, Brumme, Bunau, Nardelli, Calandra, Car, Cavazzoni, Ceresoli,
			Cococcioni et~al.}}]{giannozzi2017advanced}
	\bibinfo{author}{\bibfnamefont{P.}~\bibnamefont{Giannozzi}},
	\bibinfo{author}{\bibfnamefont{O.}~\bibnamefont{Andreussi}},
	\bibinfo{author}{\bibfnamefont{T.}~\bibnamefont{Brumme}},
	\bibinfo{author}{\bibfnamefont{O.}~\bibnamefont{Bunau}},
	\bibinfo{author}{\bibfnamefont{M.~B.} \bibnamefont{Nardelli}},
	\bibinfo{author}{\bibfnamefont{M.}~\bibnamefont{Calandra}},
	\bibinfo{author}{\bibfnamefont{R.}~\bibnamefont{Car}},
	\bibinfo{author}{\bibfnamefont{C.}~\bibnamefont{Cavazzoni}},
	\bibinfo{author}{\bibfnamefont{D.}~\bibnamefont{Ceresoli}},
	\bibinfo{author}{\bibfnamefont{M.}~\bibnamefont{Cococcioni}},
	\bibnamefont{et~al.}, \bibinfo{journal}{J. Phys.:Condens. Matter}
	\textbf{\bibinfo{volume}{29}}, \bibinfo{pages}{465901}
	(\bibinfo{year}{2017}).
	
	\bibitem[{\citenamefont{Perdew et~al.}(2008)\citenamefont{Perdew, Ruzsinszky,
			Csonka, Vydrov, Scuseria, Constantin, Zhou, and Burke}}]{perdew2008restoring}
	\bibinfo{author}{\bibfnamefont{J.~P.} \bibnamefont{Perdew}},
	\bibinfo{author}{\bibfnamefont{A.}~\bibnamefont{Ruzsinszky}},
	\bibinfo{author}{\bibfnamefont{G.~I.} \bibnamefont{Csonka}},
	\bibinfo{author}{\bibfnamefont{O.~A.} \bibnamefont{Vydrov}},
	\bibinfo{author}{\bibfnamefont{G.~E.} \bibnamefont{Scuseria}},
	\bibinfo{author}{\bibfnamefont{L.~A.} \bibnamefont{Constantin}},
	\bibinfo{author}{\bibfnamefont{X.}~\bibnamefont{Zhou}}, \bibnamefont{and}
	\bibinfo{author}{\bibfnamefont{K.}~\bibnamefont{Burke}},
	\textbf{\bibinfo{volume}{100}}, \bibinfo{pages}{136406}
	(\bibinfo{year}{2008}).
	
\end{thebibliography}



\end{document}